\documentclass[12pt,twoside]{article}

\usepackage{geometry}
\usepackage{setspace} 
\usepackage{ucs}  
\usepackage[utf8x]{inputenc}

\usepackage[T1]{fontenc}

\usepackage[german,english]{babel}

\usepackage{latexsym}

\usepackage[%
     authorformat=year,%
     titleformat=italic,%
     titleformat=commasep,%
     commabeforerest,%
     ibidem=strict,%
     citefull=first,%
     pages=format,%
]{jurabib}

\usepackage{graphicx}

\newcommand{\kopfmarke}{{\bf TEST}}
\newcommand{\autor}{Autor}
\newcommand{\titel}{Titel}
\newcommand{\institution}{Institution}
\newcommand{\summary}{Abstract}
\newcommand{\articleid}{Artikelidentifikation}
\newcommand{\wwwarticleid}{www.Adresse des Artikels.edu}

\newcommand{\key}{keytest}

\newcommand{\identitaet}[7]{
\thispagestyle{empty}

\renewcommand{\titel}{#1}
\renewcommand{\autor}{#3}
\renewcommand{\institution}{#4}
\renewcommand{\kopfmarke}{{#2}}
\renewcommand{\summary}{#5}
\renewcommand{\articleid}{#6}
\renewcommand{\key}{#7}

\mbox{ }\\
\begin{minipage}{14cm}
  
    \mbox{ }\\
    {\Large \bf \titel}\\
    \mbox{}\\
    \normalsize {\bf \autor}\\
    \normalsize (\institution)\\

    \end{minipage} \hfill

\mbox{ }\\ 
\mbox{ }\\
\textbf{\sc\ Abstract: }{\sf \summary}\\
\mbox{}\\
\textbf{\sc Keywords: }{\small\sf \key}\\
 \vspace{5mm}

}

\usepackage[pagebackref,hypertexnames=true,colorlinks=true]{hyperref}

\begin{document}  
\normalsize
\newcommand{\dbb}{de$\,$Broglie-Bohm theory}
\newcommand{\db}{de$\,$Broglie}
\newcommand{\qeh}{quantum equilibrium hypothesis}
\newcommand{\qm}{quantum mechanics}

\identitaet{What you always wanted to know about Bohmian mechanics but were afraid 
to ask\footnote{Invited talk at the spring meeting of the  {\em Deutsche Physikalische 
Gesellschaft}, Dortmund, 28.-30.3.2006}}{Introduction to the \dbb\ and QFT generalizations}
{Oliver Passon}
{Zentralinstitut f\"ur Angewandte Mathematik\\
Forschungszentrum J\"ulich\\
email: O.Passon@fz-juelich.de}
{Bohmian mechanics is an alternative interpretation of quantum mechanics. We outline 
the main characteristics of its non-relativistic formulation. Most notably 
it does provide a simple solution to the infamous measurement problem of quantum
mechanics. Presumably the most common objection against Bohmian mechanics is based on its 
non-locality and its apparent conflict with relativity and quantum field theory. 
However, several models for a quantum field theoretical generalization do exist. 
We give a non-technical account of some of these models.}
{You need not to fill this slot}
{Bohmian mechanics, \dbb , Interpretation of quantum mechanics, quantum field theory, 
theory generalization}

\section{Introduction}
This note reviews Bohmian mechanics, an alternative interpretation (or modification) 
of quantum mechanics. Bohmian mechanics reproduces all predictions of quantum mechanics
but introduces a radically different perception of the underlying processes. 
Like most alternative interpretations it is not distinguishable from standard 
\qm\ by e.g. any experimentum crucis.

We start out by a few historical remarks in Sec.~\ref{hist} before we outline the main 
characteristics of its non-relativistic formulation in Sec.~\ref{nrf}.
Here we put special emphasis on the status of ``observables'' other than
position.  However, the most important feature of the theory is its solution
to the infamous measurement problem of quantum mechanics (see Sec.~\ref{mp}).  

We then turn to the question of relativistic and quantum field theoretical
generalizations of the theory. Several such generalizations do exist and in
Sec.~\ref{gen} we give a non-technical account of some of these models.
We also address the question of what it actually means to ``generalize'' a theory
and make a little digression to the field of ``intertheory relations''.

However, before we get started, we would like to make some general remarks
concerning the interpretation of \qm . These may help to put the debate on
Bohmian mechanics into a wider context.  

\subsection{Reflections on the interpretation of \qm \label{ref}}
The interpretation of \qm\ has been discussed ad nauseam and the engagement  
with it can be a frustrating and disappointing business. This subject matter
continues to produce an endless stream of publications\footnote{\citep{cabello:2000}
  gives a bibliographic guide to the foundation of \qm\ (and quantum
  information) and collects more than 10$^5$ entries.} and nobody can
reasonably expect this issue to be settled in the future. So much the worse,
the different camps stand in fierce opposition and one gets the impression
that this is an other obstacle for reaching substantial progress. 

However, what do we actually mean by ``progress''? Perhaps, in a situation like this, 
we need to reconsider our criteria and standards for progress and success. 
   Given that the foundation of \qm\ has a smooth transition to philosophy we may learn something 
from a similar debate there. 

Chapter 15 of Bertrand Russell's little book {\em The Problems of Philosophy} (1912) 
is titled {\em The Value of Philosophy} and starts with a remark which applies just as well
to the interpretation of \qm :
\begin{quote}
``[W]hat is the value of philosophy and why it ought to be studied. It is the more necessary 
to consider this question, in view of the fact that many men, under the influence of science 
or of practical affairs, are inclined to doubt whether philosophy is anything better than 
innocent but useless trifling, hair-splitting distinctions, and controversies on matters 
concerning which knowledge is impossible.''
\end{quote}
And indeed, many practically minded physicists regard the interpretation of
\qm\ as pointless since no direct applications follow from it.  

Russell continues, that although philosophy does aim at ``knowledge which gives
unity and system to the body of the sciences'', it admittedly had little
success in this respect and could only answer very few of its questions
definitely. However, more important than the answers are the questions it
asks: 
\begin{quote}
``Philosophy is to be studied, not for the sake of any 
definite answers to its questions since no definite answers can, as a rule, be known to be 
true, but rather for the sake of the questions themselves; because these questions enlarge 
our conception of what is possible, enrich our intellectual imagination and diminish the 
dogmatic assurance which closes the mind against speculation (...)'' 
\end{quote}
Now, rated by this measure, the debate on the interpretation of \qm\ is a story of 
spectacular success indeed. Agreed, only few questions have been settled ultimately, 
but every alternative interpretation enlarges ``our conception of what is possible''.\footnote{The 
above-mentioned should not be misconceived as a license for arbitrary speculations. The possible 
answers still have to come under scrutiny.} And this is exactly what Bohmian mechanics does as 
well. It enriches our conception of what the quantum world may be.

\section{Some history \label{hist}}
Bohmian mechanics was first developed by Louis de Broglie! Therefore we will use 
the name ``\dbb '' in the remainder of this paper. Some basic concepts of the theory 
were already anticipated in de Broglie's dissertation in 1924 and his talk on the 
5th Solvay meeting in October 1927 contained an almost complete exposition of the 
theory -- called the ``pilot wave theory'' (th\'eorie de l'onde pilote) by him 
\citep{bacciagaluppi:2006}. For reasons which are not entirely clarified 
yet the theory fell into oblivion until David Bohm developed it independently in 
1951 \citep{bohm:1952}. However, the reception of this work was unfriendly, to say the least. 
See e.g. \citet{myrvold:2003} for the early objections against the \dbb . 

Since the 70s John Bell was one of the very few prominent physicists who stood up 
for the theory. Many papers in his anthology \citep{bell:2004} use the \dbb\ 
and the stochastic collapse model by \citet{ghirardi:1986} as an 
illustration of how to overcome the conceptual problems of quantum theory. 
The \dbb\ is even closely related to Bell's most important discovery, the Bell 
inequality. It was the non-locality of the \dbb\ which inspired him to develop 
this result.

Interestingly, during the 60s and most of the 70s even Bohm himself had only
little interest in his theory.  Only since the late 70s he and his group (B. Hiley, 
Ch. Dewdney, P. Holland, A. Kyprianidis, Ch. Philippidis and others) at Birkbeck 
College in London started to work on that field again. They referred to the theory 
as ``ontological'' or ``causal'' interpretation of \qm . Since the 1990th some
new groups and researchers joined the field (D. D\"urr, S. Goldstein and N. Zanghi,
A. Valentini, G. Gr\"ubl and others) and it came to the 
formation of different schools. \citet{duerr:1992} coined the 
term ``Bohmian mechanics'' which stands for a specific reading of the
theory. While mathematically equivalent to Bohm's exposition in 1952, it is influenced
by Bell's (and also de Broglie's) presentation of the theory (e.g. it puts no emphasis 
on the ``quantum potential''\footnote{It should be noted
  that while all of the before mentioned Bohm students use the quantum
  potential formulation, the presentation of the theory in \citet{bohm:1993} and 
  \citet{holland:1993} shows differences nevertheless. In addition changed
  also Bohm's own interpretation of the theory in the course of
  time. However, this is clearly not unusual and by no means specific to the
  \dbb . We just mention this point here to call into attention that -- given these
  different readings of the theory -- talking about {\em the} ``\dbb '' may
  need further qualification.}).   

Researchers who want to stay away from this debate (or who entertain their own sub-variant) 
are usually identified by calling the theory ``\dbb '', ``de Broglie-Bohm pilot wave model''
or any similar permutation of the key words.

\section{The non-relativistic formulation \label{nrf}}
The key idea of the (non-relativistic) \dbb\ \citep{debroglie:1927,bohm:1952} is
to describe a physical system not by the wavefunction, $\psi$, alone but by the couple
of wavefunction {\em and} configuration, i.e. the  position, $Q_i$, of the 
corresponding objects (e.g. electrons, atoms, or even macroscopic entities). 
\begin{center}
\fbox{\parbox{10cm}{ 
\begin{eqnarray*}
\psi &\to& (\psi, Q_i)\\
\mbox{\qm}  &\to& \mbox{\dbb}
\end{eqnarray*}
}}
\end{center}
The theory is now defined by three postulates which will be explained in the 
following\footnote{More detailed expositions of the \dbb\ can be found in 
\citet{holland:1993,bohm:1993,cushing:1994,duerr:2001,passon:2004,goldstein:2006}.}: 
\begin{enumerate}
\item  The wavefunction satisfies the usual  Schr\"odinger equation
\begin{eqnarray*}
ih\frac{\partial \psi}{\partial t}= H \psi
\end{eqnarray*}
\item  The particle velocities (a real vector field on configuration space) are given by the 
so-called {\em guidance equation}:
\begin{eqnarray}  
\label{ge}
\frac{dQ_k}{dt} &=&\frac{\nabla_k S(Q(t))}{m_k} 
\end{eqnarray}
With $Q(t)=(Q_1(t),\cdots,Q_N(t))$ the configuration of the system, $m_k$
denotes the mass of particle $k$, $\nabla_k$ is the nabla operator applied to its 
coordinates and $S$ the phase of the wavefunction in the polar representation 
$\psi=Re^{\frac{i}{\hbar}S}$. 

\item The position-distribution, $\rho$, of an ensemble of systems which are described by the
wavefunction, $\psi$, is given by $\rho=|\psi|^2$. 
This postulate is called the {\em \qeh}. 
\end{enumerate}

{\bf Postulate 1} shows that ordinary \qm\ is embedded in the \dbb\ and that
everything which is known about solutions of the Schr\"odinger equation
remains valid and important. The \dbb\ is sometimes called 
a ``hidden variable'' theory since it supplements \qm\  with additional variables, i.e. 
the particle positions. However, this terminology is a bit awkward since the positions 
are not really ``hidden''.

{\bf Postulate 2} equips the particles with a dynamic which depends on the wavefunction. 
Metaphorically speaking the quantum particles are ``riding'' on (or guided by) the 
$\psi$-field.
Thus the particles are moving on continuous trajectories and possess a well 
defined position at every instant. The proof for global existence of the Bohmian 
trajectories is given by \citet{berndl:1995} and was later extended by \citet{teufel:2005}.

The form of the guidance equation can be easily motivated.\footnote{However, its form 
is not unique. One can add an arbitrary divergence-free vector-field and arrive at 
the same statistical predictions \citep{deotto:1998}.} One may take the 
classical relation between velocity ($v$), current ($j$) and density ($\rho$): 
\begin{eqnarray}
\label{strom}
v=\frac{j}{\rho}
\end{eqnarray} 
and inserts the quantum mechanical probability current, $j$, and the probability 
density $\rho$:
\begin{eqnarray*}
j&=&\frac{\hbar}{2m_k i}\left[ \psi^* (\nabla_k \psi) - (\nabla_k \psi^*) \psi \right]\\
\rho&=&|\psi|^2 \quad .
\end{eqnarray*} 
A different motivation of the guidance equation -- based on symmetry arguments -- 
is given in \citet{duerr:1992}. 

The above equation applies only to spinless particles. However, the generalization to 
fermions (or arbitrary spin) is straightforward. One only needs to consider solutions 
of the Pauli equation 
$(\psi_1 , \psi_2)^t$ and arrives at the guidance equation \ref{strom}
with the modified current:
\begin{eqnarray*}
j=\sum_a \left ( \frac{\hbar}{2mi}(\psi_a^*\nabla \psi_a-\psi_a\nabla\psi_a^*)-
\frac{e}{mc} A \psi_a^*\psi_a\right )
\end{eqnarray*}

{\bf Postulate 3} is needed for the \dbb\ to reproduce all predictions of \qm . The 
continuity  equation of \qm\ ($\frac{\partial |\psi|^2 }{\partial t}+\nabla \left 
( |\psi|^2 \cdot \frac{\nabla S}{m} \right ) = 0$) ensures that any system will stay 
$|\psi|^2$ distributed if the \qeh\ holds 
initially.  The \qeh\ provides the initial conditions for the guidance 
equation which make the \dbb\ to obey Born's rule in terms of position
distributions. Since all measurements can be expressed in terms of position
(e.g. pointer positions) this amounts to full accordance with 
all predictions of ordinary quantum mechanics.  

Further more, the \qeh\ ensures that the \dbb\  does not allow for an experimental 
violation of Heisenberg's uncertainty principle notwithstanding the well defined 
position the particles possess in principle  \citep{valentini:1991}.

However, while it is ensured that the \qeh\ is satisfied for a configuration which is $|\psi|^2$ 
distributed once, it is by no means clear why any configuration should be accordingly distributed 
initially. {At first this seems like a very specific requirement which needs
e.g. very special initial condition of the universe. 
If the problem is viewed this way, it would be more appealing to have a
dynamical mechanism which explains why  $\rho \not= |\psi|^2$ distributed
systems {\em evolve} into a quantum-equilibrium distributed configuration. This
approach is explored in \citet{valentini:1991,valentini:1992} who claims that the  
dynamics of the \dbb\ gives rise to a relaxation into an approximate 
(i.e. coarse grained) equilibrium distribution for an enlarged set of initial
configurations. However, there exists a more convincing approach to justify the \qeh . 
Work by \citet{duerr:1992} shows, that
the \qeh\ follows by the law of large numbers from the assumption that the
initial configuration of the universe is ``typical'' for the $|\Psi|^2$
distribution (with $\Psi$ being the wavefunction of the universe). This
derivation resembles the way Maxwell's velocity distribution for a classical
gas follows from the ``typicality'' of the phase-space configuration of the
corresponding gas \citep{duerr:2004}.} 
According to this view the \qeh\ is no postulate of the \dbb\ but can 
be derived from it.\footnote{At the risk of being imprecise we gave only a short sketch of the
different strategies to motivate the \qeh . For details the reader is referred to the original 
literature.} 

\subsection{A remark on the quantum potential}
While the above presentation introduced the guidance equation as fundamental, the original work of 
\citet{bohm:1952} (and later also e.g. \citet{holland:1993} introduced the notion of a 
``quantum potential''. For the phase of the wavefunction the following equation holds: 
\begin{eqnarray}
-\frac{\partial S}{\partial t}=\frac{(\nabla S)^2}{2m} +V -\frac{h^2 \nabla^2 R}{2m R} \quad .
\end{eqnarray} 
Due to the similarity with the classical Hamilton-Jacobi equation (for the action $S$) the term 
$\propto \hbar^2$ has been baptized  ``quantum potential''. Within the Hamilton-Jacobi
theory the particle velocity is constraint to $m\cdot v=\nabla S$, which corresponds to
the guidance equation of the \dbb . 
If one adopts the quantum potential formulation the motion along the Bohmian trajectories 
can be thought of as taking place under the action of a novel ``quantum-force''. 

However, the guidance equation can be motivated e.g. by symmetry arguments \citep{duerr:1992} 
and needs no recourse to the  Hamilton-Jacobi theory. Moreover, in \citet{goldstein:1996} it is argued 
that the quantum potential formulation is misleading since it suggests that 
the \dbb\ is just classical mechanics with an additional potential (or force) term. 
But the \dbb\ is a first-order theory (i.e. the velocity is constrained by the position already)
and this important trait is disguised in the quantum potential formulation. 
 
Whether this ambiguity in the formulation of the \dbb\ should be viewed as a substantial debate
or a secondary matter depends on the context. These two readings of the theory have  
certainly a great deal in common and in comparing the de Broglie-Bohm approach with
standard quantum mechanics the distinction between these different schools is usually 
irrelevant. However, more detailed discussions which involve subtleties regarding e.g. 
the status of the wavefunction or particle properties have to pay attention to these differences.
  
\subsection{Characteristic features}
After the definition of the theory we want to discuss some of its
characteristic features and try to put them into the wider context.

\subsubsection*{Determinism}
The \dbb\ is deterministic since the wavefunction and the configuration at a 
given time fix the time evolution of the system uniquely. However, given the \qeh\ the 
predictive power of the theory is not enlarged compared to ordinary \qm . All predictions 
of the theory remain probabilistic but in contrast to ordinary \qm , the randomness is 
arising from averaging over ignorance.

However, it should be noted that to many adherents of the \dbb , determinism is not the key 
feature of the theory. For example \citet{bohm:1954} have developed a hidden variable 
model which contains a stochastic background-field and in a later section we will discuss a 
field-theoretical generalization of the \dbb\ which also contains stochastic effects. 
Moreover do many ``Bohmians'' appreciate the GRW model which includes a stochastic term into the 
Schr\"odinger equation to describe the wavefunction collapse. Short but to the point: not the 
indeterminism of \qm\ but rather its vague account of the measurement process created discomfort 
with the ordinary formulation and inspired the development of these alternative models. 

\subsubsection*{``Complementarity'' dispensable}

Many quantum phenomena (e.g. interference effects) need both, the wave and particle 
aspect of matter for their explanation. 
The notion of ``complementarity'' was developed as an attempt to justify this common use of
mutually contradictory concepts. Within the \dbb\ matter is described by a wave-like 
quantity (the wavefunction) {\em and} a particle-like quantity (the position). Hence, the 
notion of complementarity is not needed.  

\subsubsection*{Non-locality}
Since the  wavefunction is defined on the configuration space, the guidance equation 
of a $N$-particle system links the motion of every particle to the positions of the 
other particles at the same time. 
In principle the particles can influence each other over arbitrary distances. 
However, this non-locality is needed in order to explain the violation of Bell's 
inequality. Moreover ensures the \qeh\ that the correlation of space-like 
separated particles can not be used for faster than light communication 
\citep{valentini:1991}. Finally does the non-locality of the \dbb\ vanishes if the state is 
not entangled. 

Whether this non-locality is viewed as an unacceptable feature 
depends on the attitude towards the problem of non-locality in quantum mechanics in general. 
Following the work of Bell and the experimental confirmation of quantum mechanics in tests 
of the Bell inequality it became widely accepted that quantum mechanics itself is ``non-local''. 
However, the precise meaning of the term ``non-local'' is far from being unique and their 
exists a vast literature on that topic. A thorough discussion of that issue is far beyond 
the scope of the present paper (see e.g. \citet{cushing:1987}). However, one can reasonably state, 
that the ``non-locality'' of the \dbb\ is more explicit (i.e. dynamical) than the 
``non-separability'' of ordinary quantum mechanics. 

Be that as it may, given that the \dbb\ is a reformulation of non-relativistic \qm , 
any action-at-a-distance should be no threat anyway. It is turned into an objection against 
the theory if one argues that no ``Bohm-like'' relativistic or quantum field theoretical 
generalization of the theory can be given. In Sec.~\ref{gen} we will discuss the existing 
models for such generalizations.

\subsubsection*{``Measurements'' deserve no special role}
The main merit of the \dbb\ is its solution to the measurement problem. This theory  
treats ``measurements'' like any other interactions or experiments. This allows a reply 
to the frequent complaint that the trajectories of the \dbb\ violate the rule  ``Entia non sunt 
multiplicanda praeter necessitatem'' which is usually attributed to William of Ockham 
(``Ockham's razor''). While the trajectories are additional entities indeed, any  
``measurement postulate'' or the like becomes unnecessary. Given the importance of this
point we devote Section~\ref{mp} to a more detailed discussion of the measurement-problem 
and how it is solved by the \dbb .

\subsubsection*{``Observables'' other than position and contextuality} 
Much more important than being deterministic or having particle trajectories is the   
novelty of the \dbb\ with regard to the status of ``observables'' other than position.
Within ordinary \qm\ the identification of ``observables'' with linear Hilbert space operators 
is usually regarded as the key innovation. Their non-commutativity is believed to be the 
mathematical embodiment of the deep epistemological lesson \qm\ teaches us. 

The \dbb\ takes a different route. First, it includes the particle positions (which are 
described by real coordinates, and not by some operator) into the state description.
Second, it distinguishes these variables, i.e. the outcome of every experiment is determined by 
the wavefunction and the configuration. Note, that this holds also for experiments which 
are supposed to ``measure'' quantities like  energy, angular momentum, spin etc.
There are no ``hidden variables'' or continuous functions which correspond to the ``actual'' 
values of these quantities\footnote{In fact, \citet{holland:1993} p.~91ff, introduces  ``local 
expectation values'' for these quantities which are supposed to correspond to their ``actual'' 
value along the trajectories. Averaged over the quantum equilibrium distribution these local 
expectation values reproduce the quantum mechanical predictions. However, one might object 
that these ``properties'' are redundant since the position is
already enough to reproduce all experimental predictions of \qm . Further more they
are not conserved along the Bohmian trajectories.}. Within the \dbb\ all these 
quantities do have a different ontological status than position.  D\"urr {\em et al.} write 
(using spin as an example only):
\begin{quote}
``Unlike position, spin is not {\em primitive}, i.e., no {\em actual} discrete degree of 
freedom, analogous to the {\em actual} positions of the particles, added to the state 
description in order to deal with ``particles with spin''. Roughly speaking, spin is 
{\em merely} in the wave function.'' (\citet{duerr:1996}, p.11)
\end{quote}
In common jargon these properties are called ``contextual'', i.e. the measurement
does not reveal a pre-existing value of a system-property but depends crucially on 
the experimental arrangement (the ``context'').\footnote{In \citet{duerr:2004}, p.64ff, 
it is argued that the term ``contextual property'' is actually 
misleading because it suggests that e.g. spin is still a ``property''.  But 
``properties which are merely contextual are no properties at all'' (\citet{duerr:2004}, p.67).} 

Thus, in general, ``measurements'' do not measure anything in the closer meaning of the 
term. The only exception being of course position measurements, and, in some sense 
momentum-measurements. The latter do indeed measure the asymptotic (Bohmian) velocities. 
Hence, the only properties of a ``Bohmian particle'' are its position and its velocity. 
Just as $\psi$ is no classical field, the Bohmian particles are  no classical particles, 
i.e. they are no bearers of properties other than position. {Therefore a physical object 
like e.g. an electron should not be confused with the Bohmian particle at position $Q_i$. It is 
represented by the {\em pair} $(\psi, Q_i)$. 

Agreed, this is a radical departure from
the classical particle concept. However, within the \dbb\ this move is not only natural
(recall that e.g. momentum and energy are concepts which arise in 2nd order Newtonian 
mechanics while the guidance equation of the \dbb\ is 1st order)
but allows for an elegant circumvention of the  Kochen-Specker ``no-go'' 
theorem, directed against hidden variable theories (see e.g. \citet{mermin:1990}.}
This theorem demonstrates, that a consistent assignment of possessed
values to all observables for a quantum mechanical state is not possible. 
However, if  you allow for contextuality -- as the \dbb\ does -- you do not 
{\em expect} such an assignment to exist at all. 



According to \citet{duerr:2004} the ``naive realism about operators'', i.e.
the identification of operators with properties and the common talk about ``measuring'' 
operators, is the source of most of the confusion in the interpretation of \qm .
However, given what we have said above, it {  may appear} puzzling why operators can play 
such a prominent role in the usual formulation of \qm\ and how exactly they relate
to the Bohmian formulation. In  \citet{duerr:2004} it is shown how operators
naturally arise in the \dbb . They are derived quantities which are coding the probability 
distributions for certain ``measurement-like'' (p.11) experiments. This leads us to 
the next section which is devoted to a discussion of how the \dbb\ treats ``measurements'' 
and in particular how it solves the measurement problem.

\subsection{How the \dbb\ solves the measurement problem \label{mp}}
Let us first briefly recall the measurement problem of \qm . It can be stated in several ways, 
e.g. \citet{maudlin:1995}, p.7, offers the following formulation:\footnote{In fact, 
\citet{maudlin:1995} introduces three slightly different formulations of the measurement problem. 
We refer only to the first formulation (hence, Maudlin labels the following propositions 1.A, 
1.B and 1.C).}
\begin{quote}
The following three claims are mutually inconsistent:

A The wave-function of a system is {\em complete}, i.e. the wave-function specifies   
(directly or indirectly) all of the physical properties of a system.

B The wave-function always evolves in accord with a linear dynamical equation (e.g. 
the Schr\"odinger equation).

C Measurements of, e.g. the spin of an electron always (or at least usually) have 
determinate outcomes [...]
\end{quote}
The argument runs like this: Given a two-valued observable $S$ 
with eigenvectors $\psi_1$ and $\psi_2$.  Let  $\Phi_0$ denote its wavefunction in 
the ``ready-state'' and  $\Phi_1$ ($\Phi_2$) the state of the apparatus if the measurement 
yields  $\psi_1$ ($\psi_2$). Hence, $\hat{U}(\psi_i \otimes \Phi_0) = \psi_i \otimes \Phi_i$ 
($i\in \{1,2\}$) holds,  with $\hat{U}$ the time evolution of the combined system. A general 
state will be a superposition:
\begin{eqnarray*} 
\psi=c_1\psi_1 + c_2\psi_2 
\end{eqnarray*}
Now, given B, the action of  $\hat{U}$ on this state yields:
\begin{eqnarray} 
\label{cat_state}
\hat{U}(\psi \otimes \Phi_0) = c_1\psi_1\otimes\Phi_1 + c_2\psi_2\otimes\Phi_2 
\end{eqnarray} 
While individual measurements always result in {\em either} the state $\Phi_1$ {\em or}  
$\Phi_2$, this is a superposition of different pointer states. Thus, in contrast to our 
experience quantum mechanics does not leave the joint object-apparatus system  
in a definite state.\footnote{Our argument relied on simplifying assumption
  like  an ideal measurement and pure states for both, object and 
  apparatus. One might suspect that the problem is only generated by these 
  unrealistic conditions. However, even in the completely general case
  employing density operators (i.e. mixed states), non-ideal measurements,
  interactions with the environment etc.pp. the conclusion remains essentially
  unaltered (see \citet{bassi:2000} and \citet{gruebl:2003}).}
According to assumption A the wave-function should specify every physical fact 
about the measurement device. Maudlin argues that, since the two $\Phi_i$ enter 
symmetrically, it is  not clear by what argument one could attempt to show that the 
final state \ref{cat_state} represents one but not the other indicator state. Thus, 
assuming A and B contradicts C. Any resolution of this problem has to deny at least one of the 
above assumptions. 

To deny proposition A needs some sort of ``hidden'' (or actually ``additional'') variables. 
The \dbb\ is a prominent example for this strategy and we explain how this solves the 
measurement problem further below. {Ballentine's  statistical or ensemble interpretation 
\citep{ballentine:1970} can also be construed as a denial of proposition A. It takes  the quantum 
state to be the description of the statistical properties of an ensemble of identically 
prepared objects only.} 

To deny proposition B leads to so-called ``collapse theories'' which abandon the strict 
linear time evolution of the system. For example \citet{ghirardi:1986} have 
developed 
such a non-linear model which describes this mechanism. Also does von Neumann's proposal 
of a collapse of the wavefunction fall into this category. However, von Neumann (like all 
other standard presentations of \qm) did not specify the physical conditions under which 
the linear evolution fails.  

Finally one may question C and the many-world interpretation can be construed as a solution of 
the measurement problem along this line.

\subsubsection*{Effective collapse in the \dbb}
Now we turn in more detail to the \dbb\ and its resolution of the measurement problem. 
It denies assumption A from the previous section, i.e. introduces the particle position as 
{\em additional} variables to arrive 
at a complete state description. However, what is needed are not just ``additional'' 
variables but variables which supply the necessary means to distinguish different 
measurement outcomes.\footnote{\citet{maudlin:1995}, p.11, notes that therefore ``additional'' 
variables which would be really ``hidden'' (i.e. unobservable) would not help at all.} 

Quantum mechanics describes how a superposition state evolves into a sum of 
macroscopic distinct (i.e. non-overlapping) states, i.e. $(\psi_1\otimes\Phi_1) \cdot 
(\psi_2\otimes\Phi_2) \approx 0$. It just fails to distinguish the branch which
corresponds to the actual measurement outcome. Within the \dbb\ the different measurement 
outcomes correspond to different {\em configurations} (e.g. pointer positions). The 
positions provide a record of the measurement outcome, or more generally they ``yield 
an image of the everyday classical world'' (\citet{bell:2001}, p.41). 

Suppose for example that the measurement yields outcome ``1'', i.e. the initial position
of the Bohm particle was such that the deterministic evolution developed into a configuration
that lies within the support of $\psi_1\otimes\Phi_1$. The Bohm particles will be guided by 
this state because the non-overlapping $\psi_2\otimes\Phi_2$-part is dynamically 
irrelevant. Thus the \dbb\ provides a so-called ``effective collapse'' of the wavefunction. 
Given the \qeh\ the probability for this effective collapse obeys Born's rule.

\section{Relativistic and quantum field theoretical generalizations \label{gen}}
Presumably the most common objection\footnote{A comprehensive discussion of objections 
against the \dbb\ can be found in \citet{passon:2004b}} against the \dbb\ is based 
on its non-locality and its apparent conflict with relativity and quantum field theory. 
However, several ``Bohm-like'' models for relativistic quantum mechanics and quantum 
field theory do exist. Here we give a non-technical account of some of these models.
But before doing so, we need to say a few words on the actual meaning of ``Bohm-like''. 

\subsection{What is a ``Bohm-like'' theory? \label{what}}
At first sight ``Bohm-like'' seems to mean ``having trajectories'' or even ``having 
deterministic trajectories''. Obviously this requirement is intended to capture the 
spirit of the \dbb . The task of developing e.g. a Bohm-like quantum field 
theory is then to reconcile this concept with the predictions of QFT. 

This may even be possible (see for example the Bell-type models below), however, on closer 
inspection this requirement seems to be too narrow nevertheless. 
One only needs to consider the history of physics, where many important features of a given 
theory did not carry over to its generalization. In particular does QFT provides  
examples for the departure from concepts which were accepted in non-relativistic 
quantum mechanics. Or to put it differently: one should expect (or at least not exclude 
from the outset) new concepts to enter a theory if it is extended to new areas.

Another more reasonable demand for a quantum field theoretical generalization of the 
\dbb\ is that it (i) reproduces the predictions of QFT and (ii) includes the non-relativistic
formulation as a limiting case. The last requirement seems necessary to regard a model
as a generalization. In Sec.\ref{gen_gen} we will come back to this important question.   

However, the existing models for ``Bohm-like'' QFT concentrate on still another feature 
of the \dbb . They suggest, that the essence of the \dbb\ is its ``clear ontology'', i.e. 
that it attributes ``being'' to certain entities. In common jargon, the theory possesses 
``beables''. This term was coined by \citet{bell:1976} and is meant in contrast to ``observable''  
i.e. emphasizes that any observation (i.e. measurement) deserves no special role in the 
formulation of a fundamental theory. In Bell's own words: 
\begin{quote}
``In particular we will exclude the notion of ``observable'' in favor of that of ``beable''. 
The beables of the theory are those elements which might correspond to elements of reality, 
to things which exist. [...] Indeed observation and observers must be made out of beables.'' 
(\citet{bell:1986}, p.174) 
\end{quote}
The beables of the non-relativistic \dbb\ happen to be particles {  (a good 
question is whether the wavefunction $\psi$ should be regarded as a beable likewise. Bell 
regarded the state-vector as a beable, ``although not a local one'' (\citet{bell:1986}, p.176))} 
which move on continuous trajectories. In what follows we will also
come across field-beables and indeterministic dynamics in ``Bohm-like''
theories. As long as this beables provide the means to record measurement outcomes
they can be used to build a Bohm-like model.

\subsection{The Bohm-Dirac theory}
We begin with the question of a relativistic generalization. Already in \citet{bohm:1953}
an extension of the \dbb\ to the Dirac equation was given. The strategy here
is analogous to the non-relativistic case.  Solutions of the Dirac equation 
fulfill a continuity equation with a time-like current. {  The spatial part of this current reads 
$\psi^{\dagger} \alpha_k \psi$.} In addition the  density $\rho=\psi^{\dagger}\psi$ (the
appropriate {\em quantum equilibrium distribution}) is positive definite. 
Thus, similar to the non-relativistic case a particle velocity can be defined by the
ratio of these two quantities:
\begin{eqnarray}
\frac{dQ_k}{dt} &=& {{\psi^{\dagger} \alpha_k \psi}\over {\psi^{\dagger}\psi}}\\
\qquad \mbox{with:} &&  \alpha_k^i = 1\otimes \cdots \otimes \alpha^i \otimes \cdots \otimes 1
\quad \mbox{and:} \;\alpha^i=\left ( \begin{array}{cc} 0  & \sigma_i \\ \sigma_i & 0 \end{array} 
\right )  \nonumber
\end{eqnarray}
In this way the description is complemented by the configuration, i.e. the beables 
of this theory are particles as in the non-relativistic formulation.  

However, in the many-particle case this theory is not Lorentz covariant
since it uses a common time for all particles. The frame-of-reference in which
$\rho=\psi^{\dagger}\psi$ holds is distinguished \citep{berndl:1995b}. 
But this non-covariance is only relevant on the level of individual particles. The statistical
predictions of the Bohm-Dirac theory are the same as for the usual Dirac theory 
because (i) by construction it is ensured that they hold in the distinguished frame
and (ii) they transform properly under Lorentz transformations.
Hence, the preferred frame-of-reference can not be identified experimentally. 

In fact, as shown by \citet{duerr:1999}, it is even possible to formally 
restore Lorentz invariance for the Bohm-Dirac theory by introducing additional structure. 
D\"urr {\em et al.} introduce a 
preferred slicing of space-time, determined by a Lorentz invariant law. 

In order to deal with anti-particles one might invoke the Dirac-sea concept, i.e.
introduce particle beables for every negative energy state (\citet{bohm:1993}, p.276). 

Other approaches to develop a relativistic \dbb\ use the concept of the  
multi-time wavefunction $\psi(q_1, t_1, \cdots , q_N, t_N)$, i.e. introduce a different 
time variable for each particle. However, the resulting set of coupled Dirac equations 
can only be solved in the absence of interaction potentials. See \citet{tumulka:2006} and the 
references therein for a more detailed discussion of these models.

However, it is generally agreed that the unification of quantum mechanics and 
relativity needs a quantum field theoretical framework anyway. We therefore turn to the 
field theoretical generalizations of the \dbb . Here several competing models do exist.

\subsection{Quantum field theoretical generalizations}
We have learned in Sec.~\ref{what}, that the {\em beable} is the decisive quantity 
in a Bohm-like theory. Hence, the different models for a quantum field theoretical 
generalization of the \dbb\ can be classified according to the beables they employ. 
Roughly the models fall into the following three categories:

\subsubsection*{Field-beables for bosons and particle beables for fermions}
Already in his seminal paper in 1952 Bohm presented a way of generalizing his causal 
interpretation to the electromagnetic field. The additional variables 
(or beables) were not particles but fields. The quantum state is thereby a wavefunctional 
which guides the field beable. This approach can be extended to the various bosonic fields 
(see e.g. \citet{bohm:1984,holland:1993,kaloyerou:1996}. For example the second-quantized real 
Klein-Gordon field is described by a wavefunctional $\Psi(\phi({\bf x}),t)$, which satisfies 
the Schr\"odinger equation:
\begin{eqnarray} 
i { \partial \Psi \over \partial t } = \int d^3 x \left( - \frac{\delta^2 }{\delta \phi^2}  
+ (\nabla \phi)^2 \right)\Psi .
\end{eqnarray}
The corresponding guidance equation for the field beable $\phi({\bf x},t)$
reads 
\begin{eqnarray} 
 { \partial \phi \over \partial t } =   { \delta S \over \delta \phi }, 
\end{eqnarray}
where $S$ is the phase of the wavefunctional $\Psi$. 

In these models the configuration space is the infinite dimensional space of field 
configurations. Since there does not exist a Lebesgue volume measure on these spaces  
the rigorous definitions of an equivariant measure, i.e. the analogue of $|\psi(q)|^2dq$,
is problematic (\citet{tumulka:2006}, p.12).

For fermionic quantum fields Bohm {\em et al.} argue that a causal 
interpretation in terms of field beables cannot be constructed \citep{bohm:1987} 
and (\citet{bohm:1993}, p.276). Instead Bohm and Hiley propose to introduce 
particle beables for fermions according to the Bohm-Dirac theory mentioned above. 
In fact, models by Holland and Valentini which try to provide field-beables for 
fermions did not succeed (\citet{struyve:2006}, p.1).

\subsubsection*{Field-beables for bosons and no beable-status for fermions}
Inspired by the difficulties to construct a Bohm-like theory for fermions with 
field-beables, \citet{struyve:2006} 
propose a different direction. They recall that e.g. the property ``spin'' can be described 
in the \dbb\ without assigning a beable status to it. They suggest, that the same 
may be done for the fermionic degrees of freedom. Since fermions are always gauge-coupled 
to bosonic fields it is sufficient to introduce beables for the bosons. 

Technically their work is similar to Bohm's model with field-beables for bosons 
mentioned above. They introduce a specific representation for the bosonic field-operators
and trace out the fermionic degrees of freedom. Their beables are the transversal part 
of the vector potential. In \citet{struyve:2006} this approach is carried out for 
QED, but it has a natural extension to other gauge theories. 

Struyve and Westman discuss in detail how this model accounts for an effective 
collapse, i.e. how the total wavefunctional evolves to a superposition of non-overlapping 
wavefunctionals. However, one might still worry if this model is capable to contain a record 
of the measurement outcome, for example in terms of pointer positions. They reply to this 
concern, that 
\begin{quote}
``(...) if we continue our quantum description of the experiment, the direction of the 
macroscopic needle will get correlated with the radiation that is scattered off (or 
thermally emitted from, etc.) the needle. Because these states of radiation will be 
macroscopically distinct they will be non-overlapping in the configuration space of 
fields and hence the outcome of the experiment will be recorded in the field beables 
of the radiation.''(p.18)
\end{quote}

We now turn to an approach which can be viewed as complementary to the Struyve-Westman model.
While their model views fermions as an epiphenomenon, the Bell model we are going to discus next 
can be seen as tracing out the bosonic degrees of freedom (\citet{struyve:2006}, p.8).

\subsubsection*{Particle beables for fermions}

\citet{bell:1986} presented a model for Hamiltonian quantum field theories with
the  fermion number as beable. He regarded this to be a natural
generalization of the particle concept, since 
\begin{quote}
``The distribution of fermion number in the world certainly includes the
  positions of instruments, instrument pointers, ink on paper, ... and much
  much more.'' (p. 175) 
\end{quote}
Hence, to assign beable status to this quantity ensures a solution of the 
measurement problem.\footnote{However, Bell acknowledges that this beable choice is everything but 
unique (p.179).} This model is formulated on a spatial lattice with points
enumerated by $l=1,2,\cdots, L$ (the time remains continuous). For each lattice site a fermion
number operator is defined with eigenvalues $F(l)=0,1,2,\cdots , 4N$ ($N$ being 
the number of Dirac fields).  

The ``fermion number configuration'' at each time is thus the list $n(t)=(F(1),\cdots , F(L))$. 
While the non-relativistic \dbb\ regards $(\psi,Q_i)$ to be the complete specification of the state 
of a system, this model considers the pair $(|\psi \rangle , n)$ (with $|\psi \rangle $ being 
the state vector).

The task is now to find the proper dynamics for this pair. For the state vector the usual 
evolution  
\begin{eqnarray*}
\frac{d}{dt}|\psi(t) \rangle=\frac{1}{i}H |\psi(t) \rangle
\end{eqnarray*}
is considered (in the following $\hbar$ is set to 1). Again this gives rise to a continuity 
equation:
\begin{eqnarray} 
\label{ce_bell}
                        &&\frac{d}{dt}P_n = \sum_m J_{nm} \\
\mbox{with:} \qquad P_n &=& \sum_q | \langle n,q | \psi(t) \rangle |^2  \nonumber\\
                  J_{nm}&=& \sum_{q,p} 2\mathrm{Re} \langle \psi(t)|n,q \rangle \langle n,q| -iH|m,p  
                                                                                           \rangle 
\langle m,p|\psi(t) \rangle \nonumber
\end{eqnarray}
Here $q$ and $p$ denote additional quantum numbers such that e.g. $|p,n\rangle$ forms a basis in 
Hilbert space. The $n$ and $m$ in the state specification denote the fermion number.
Thus $P_n$ is the probability distribution for the fermion number configuration $n$.
While ordinary quantum mechanics (or quantum field theory) views this as the probability 
to {\em observe} the system in this state, Bell views it as the probability
for the system to {\em be} in this state. Therefore it is his ambition to
establish an analog to the guidance  equation, i.e. to describe the time
evolution of this {\em beable} irrespectively of its being observed or not. 

Bell prescribes a stochastic evolution\footnote{Bell expected the
  indeterminism to disappear in the continuum limit.} for the fermion 
number with the jump rate $T_{nm}$, i.e. the probability to jump to the 
configuration $n$ within the time span $dt$, given that the present 
configuration is $m$, is given by $T_{nm} dt$. Clearly the following equation holds: 
\begin{eqnarray}
\label{master}
\frac{dP_n}{dt}=\sum_m (T_{nm}P_m-T_{mn}P_n),
\end{eqnarray}
i.e. the change of $P_n$ in time is given by the jumps $m\to n$ diminished by  
the jumps $n\to m$. However, Equ.\ref{master} must be reconciled with condition \ref{ce_bell}, 
i.e. the stochastic dynamics needs to obey the continuity constraint. This leads to the 
condition $J_{nm}=T_{nm}P_m-T_{mn}P_n$, which is for example  satisfied by the 
choice:\footnote{This choice is not unique, e.g. one may add solutions of the homogeneous equation.}
\begin{eqnarray*}
T_{nm} = \left \{ \begin{array}{r}  J_{nm}/P_m \quad \mbox{if} \quad J_{nm}>0\\  
                                          0  \quad \mbox{if} \quad J_{nm}\le 0  
                \end{array} \right .
\end{eqnarray*}
Finally, the probability $T_{nn}dt$ for the system to remain in the same 
fermion number configuration is fixed by the normalization $\sum_m T_{mn}dt=1$.  
Given an initial configuration of the fermion number in accordance with  
$P_n(t_0)=\sum_q | \langle n,q | \psi(t_0) \rangle |^2$ this model reproduces 
all predictions of ordinary quantum field theory.\footnote{Bell notes that this includes also 
the outcome of the Michelson-Morley experiment, although this formulation relies on a particular 
division of space-time. Hence the violation of Lorentz invariance is not detectable (p.179).} 

The physical picture is that the world describes a random walk in the fermion-number 
configuration space; this random walk being biased by the state $|\psi(t) \rangle$.  
The non-deterministic jump processes correspond to the creation and annihilation of 
particles.

\citet{duerr:2004b,duerr:2005} have developed a similar process in the continuum for more or less 
any regularized quantum field theory and call it ``Bell-type quantum field theories''. While 
their model is continuous it still includes a random processes i.e. is non-deterministic. 
However, work of Colin (2003) suggests that it is also possible to construct a deterministic 
continuum limit. The difference between these two continuum versions of the Bell-model lies
in the treatment of the vacuum. D\"urr {\em et al.} take it to be the state 
with no particle-beables. In contrast does Colin's model introduce particle beables for 
every negative energy solution, i.e. invokes the Dirac sea concept. Thereby the configuration 
space becomes infinite dimensional, i.e. does not possess a Lebesgue volume measure. As mentioned 
before in the context of field-beables this introduces problems for a rigorous definition of an 
{\em equivariant measure} (\citet{tumulka:2006}, p.15).

\subsection{Some remarks on theory-generalization \label{gen_gen}}
In Sec.\ref{what} we have argued that having beables qualifies a theory as
``Bohm-like''. Further more we have used the expression ``Bohm-like'' and 
``generalization of the \dbb '' synonymously. However, there seem to be reasonable
distinctions between these two concepts. In the remainder of that paper we want to
discuss the issue of theory generalization in some more detail. We will argue that
being a ``generalization of the \dbb '' is actually a more restrictive property than
being ``Bohm-like'' only. We investigate whether this may help to single out a candidate
from the competing models discussed in the previous section. However, we will also see
that this is complicated by the fact that the concept of ``theory generalization''  is 
more involved than usually considered.

\subsubsection*{Do all ``Bohm-like'' models generalize the de Broglie-Bohm theory?}
So far we have been discussing ``Bohm-like'' QFT or actually ``beable-QFT''. 
However, we have already indicated in Sec.~\ref{what}, that in order to regard these 
models as a ``generalization'' of the original theory it is reasonable to demand a specific 
relation between the non-relativistic formulation and these models. 
Very natural is the requirement that the Bohm-like QFT 
should include the non-relativistic \dbb\ as a {\em limiting case}. After all, there is no strict 
boundary between non-relativistic and relativistic physics and the corresponding theories 
should ideally merge to each other. We want to call this our preliminary criteria for 
``theory generalization''.

\citet{vink:1993}, p.1811, investigates the relation between his generalized Bell-model 
and the original \dbb . He shows that the stochastic dynamics leads to the ordinary 
\dbb\ in the continuum limit. His argument is mathematically not rigorous but given 
that this model employs a particle-ontology from the outset it is certainly plausible 
to expect such a limit to exist.

The situation seems very different when it comes to field-beables; for example in 
the Struyve-Westman model. Given that there the fermionic degrees 
of freedom have no beable status it is not conceivable how to obtain the non-relativistic 
formulation as a limiting case. One may illustrate this with the example of the hydrogen 
atom. In the \dbb\ the physical picture of this system is a particle-beable (assigned to 
the electron) distributed according to $|\psi|^2$. In the Struyve-Westman model only
the radiations degrees of freedom of the electromagnetic field have beable status and the 
``electron'' is only an epiphenomenon. Therefore the Bohm-like 
QFT  {\`a} la Struyve and Westman can not be viewed as a generalization of the ordinary \dbb\ 
(in the above sense) but provides a complete reformulation of the non-relativistic theory.

Thus, the criteria whether a Bohm-like QFT includes the \dbb\ as a limiting case
seems to allow an assessment of the different models. Rated by this measure the Bell-type
models seem to be superior since they start with the same ontology as the non-relativistic 
formulation from the outset. But do we really have compelling arguments to make the 
non-relativistic formulation the touchstone for QFT generalizations?
One could also be willing to modify the non-relativistic \dbb\ (e.g. along the lines sketched above 
in the hydrogen example). It seems reasonable to argue that not the non-relativistic formulation 
itself but only its predictions need to be recovered. 

But there is even another twist in the above argument. Sofar we have employed a 
specific  concept of ``theory generalization'' (the limiting case relation) and found that
the field-beable approach has problems to cope with it. However, one may also ask how natural 
the requirement of the limiting case relation actually is. In fact these and related 
intertheory relations have  been critically examined within the philosophy of science. 
We will therefore say a few words on this debate and its possible impact on our question.

\subsubsection*{What does it mean to ``generalize'' a theory?}
Within the philosophy of science this question is part of the study of 
{\em intertheory relations} \citep{batterman:2005} and offers some surprises. 

Traditionally this and related questions were framed in the context of ``reductive relations'' 
between theories, i.e. the question whether a given theory $T_1$ (the primary theory) reduces to 
$T_2$ (the secondary theory).\footnote{Here we take ``reduction'' to be the move from the 
general (i.e. more fundamental) to the specific. In the philosophical literature it is often
regarded the other way around.} In some sense ``theory generalization'' is the inverse operation to 
``theory reduction''. An early and influential treatment of theory reduction was given by 
\citet{nagel:1961} (Chapter 11) who viewed theory reduction essentially as a relation of deduction, 
i.e. the laws of the secondary theory should be derivable from the laws of the primary theory.
However, this typically requires a translation of the descriptive terms of $T_2$ which are absent 
in $T_1$ into the $T_1$-language (so-called ``bridge principles''). 

In reply to criticism against the highly idealized picture of the Nagelian 
account more sophisticated models of reduction have been developed (e.g. \citet{schaffner:1967,schaffner:1969,nickles:1973} and \citet{hooker:1981}). 
Our above discussion used the notion, that a theory, $T_1$, reduces to an other, $T_2$, 
if $T_2$ is obtained as a {\em limiting case}, i.e. if there is a parameter, say $\epsilon$, 
in the primary theory such that the laws of the secondary theory are obtained in the limit 
$\epsilon \to 0$. This is a modification of the Nagelian account due to  \citet{nickles:1973}. The 
textbook example is the relation between special relativity and classical mechanics in the 
limit $(v/c)^2 \to 0$. 

However, it has been shown that this notion of reduction can not account for many relevant 
cases. For example the mathematical physicists Sir Michael Berry noted with respect to this 
example, that 
\begin{quote} 
``(...) this simple state of affairs is an exceptional situation. Usually, limits of 
physical theories are not analytic: they are singular, and the emergent phenomena 
associated with reduction are contained in the singularity.'' (\citet{berry:1994}, p.599) 
\end{quote} 
In such cases there is no smooth reduction relation between the corresponding theories, i.e.
the secondary theory can neither be derived from the primary theory nor obtained as a 
limiting case, since the limit simply does not exist.\footnote{ 
A simple example of a singular limit is given by \citet{batterman:2005}. The equation $x^2\epsilon+x-9=0$ 
has two roots for any value of $\epsilon > 0$ but only one solution for the $\epsilon = 0$ case. 
Thus, the character of the behavior in the case $\epsilon=0$ differs fundamentally from the 
character of its limiting (i.e. $\epsilon$ small but finite) behavior.} 
Examples investigated by Berry are 
the relation between wave and ray optics or quantum and classical mechanics.\footnote{Interestingly 
this is not taken as evidence against reduction per se. Berry states, that ``what follows should 
not be misconstrued as antireductionist. On the contrary, I am firmly of the view [...] that all 
the sciences are compatible and that details links can be, and are being, forged between them. But 
of course the links are subtle [...]'' (\citet{berry:2001}, p.4).} In fact the classical limit of 
quantum mechanics belongs to the open foundational questions of the theory (see 
\citet{landsman:2005} for an excellent overview). 

Thus, there are many relevant cases in physics which intuitively count as ``theory generalization'' 
but fail to satisfy  the limiting-case relation. If one is not willing to loose these 
cases one can not require this condition.

With respect to the relation between higher level and lower level (i.e. more fundamental) theories 
some authors argue for a relation called ``emergence''. The different versions of emergence roughly 
share the idea that ``emergent entities (properties or substance) `arise' out of more fundamental 
entities and yet are `novel' or `irreducible' with respect to them'' \citep{oconnor:2002}. 
Another way to characterize emergence is simply by a denial of reduction (R-emergence) or a denial 
of supervenience\footnote{Supervenience may be characterized as an ontic relation between 
structures, i.e. sets of entities  together with properties and relations among them. A structure 
$S_A$ is said to supervene on an other, say $S_B$, if  the A-entities are composed of B-entities 
and the properties and relations of $S_A$ are determined by properties and relations of $S_B$. 
It should be noted that neither does reduction entails supervenience nor the other way around.}  
(S-emergence) (see \citet{howard:2003}, p.3ff).

However, if one denies the possibility to {\em reduce} a theory from a more fundamental level, 
the inverse move (i.e. the theory generalization) is affected as well. In what sense should a 
theory $T_1$ be regarded as a generalization of (i.e. being more ``fundamental'' than) a theory 
$T_2$ if it is not possible to recover $T_2$ from $T_1$? The whole talk about ``higher level'', 
``lower level'' or being ``more fundamental'' becomes void and  one seems to be left over
with autonomous theories.

These brief remarks shall indicate that the concept of a ``theory generalization'' is more 
involved than usually considered (at least in the physics community). Thus, the failure of e.g. 
Bohm-like QFT with field-beables 
to recover the ordinary \dbb\ as a limiting case may be viewed rather as a generic feature  
in the relation between ``higher'' and ``lower'' level theories and not as a reason to reject 
this model.

It might still be possible to justify a certain beable choice based on the criteria
that the relation between the corresponding Bohm-like QFT and  the non-relativistic \dbb\
has desirable properties. However, this needs a more refined definition of ``theory 
generalization''. It seems very promising to investigate the Bohm-like quantum field 
theories as case studies for intertheory relations in order to learn more about 
both, ``theory generalization'' in general and the de Broglie-Bohm-program in particular.

\section{Summary and conclusion}
The non-relativistic \dbb\ is able to give an observer independent account of
all quantum phenomena. It solves the infamous measurement problem, or, to be
more precise, there is no such problem in the \dbb . It serves as a counter
example to the common claim that no description of quantum phenomena  
can be given which employs particles moving on continuous trajectories. 
However, like most alternative interpretations it is not experimentally distinguishable 
from standard \qm .

When it comes to relativistic and quantum field theoretical generalizations one 
first needs to agree upon what one actually means by 
a ``Bohm-like'' theory. Seemingly a theory needs to have deterministic trajectories 
to count as ``Bohm-like''. However, most Bohmians would suggest that the  decisive property 
of the \dbb\ is that it attributes a ``beable-status'' to certain properties. As long 
as these beables provide the means to record measurement outcomes they can be used 
to build a Bohm-like model. Particle beables are just a specific example for this strategy. 
For relativistic and quantum field theoretical generalizations several competing models 
do exist. These display a surprising flexibility with respect to the ``beable-choice''. Some 
models stick to a particle ontology while others introduce field-beables. Further more there 
is no need to introduce beables for all particle species and e.g. the Struyve-Westman model 
does without a beable status for fermions.\footnote{The question whether all particles (should)
have beable status is also addressed in \citet{goldstein:2005}.} 

A further investigation of the relation between these different models and the original \dbb\ 
seems to be an interesting case-study for what has been called ``intertheory relations'' in the 
philosophy of science. Possibly an assessment of these models could be based on the result. 
 
Be that as it may, the common claim that the \dbb\ is incompatible with quantum field theory 
is certainly incorrect. Agreed, all these models have a ``cooked-up'' flavor, but this is due 
to the fact that their task is (in general) to reproduce the predictions of existing theories.
These existing theories work FAPP (for all practical purposes) and the ambition 
of ``Bohm-like'' reformulations is not to extend their predictive power but to put 
them on a conceptually firm basis.  

Now, does this mean that every physicist should be a Bohmian? Certainly not. But
those who reject this possible quantum world should use correct arguments.

\subsection*{Acknowledgment}
I am grateful to Brigitte Falkenburg for inviting me
to the spring meeting of the Deutsche Physikalische Gesellschaft in Dortmund
and thank Roderich Tumulka and Ward Struyve for valuable comments to this paper.


\bibliographystyle{jurabib}
\bibliography{bohm_final_arxiv}

\end{document}